\documentstyle[12pt]{article}

\newcommand{\bcn}{\begin{center}}
\newcommand{\beq}{\begin{equation}}
\newcommand{\beqn}{\begin{eqnarray}}
\newcommand{\ecn}{\end{center}}
\newcommand{\eeq}{\end{equation}}
\newcommand{\eeqn}{\end{eqnarray}}

\newcommand{\sect}[2]{\vspace*{6mm}\hspace*{-\parindent}{\bf #1.}~{\bf
#2}\vspace*{4mm}}

 \def\lsim{\mathrel{\rlap{\lower4pt\hbox{\hskip1pt$\sim$}}
    \raise1pt\hbox{$<$}}}         

\begin{document}
\bcn
{\bf THREE SUCCESSFUL TESTS OF COLOR TRANSPARENCY AND NUCLEAR
FILTERING}\footnote {To be published in the Proceedings of the V Rencontre de
Blois meeting on Elastic and Diffractive Phenomenon, edited by C. I. Tan and H.
M. Fried (Brown University 1993) (World Scientific, in press)}\vspace*{1cm}

{\bf PANKAJ JAIN}\\

\bigskip
{\it Department of Physics and Astronomy\\
University of Oklahoma\\
Norman, OK 73019}\\

\bigskip
and
\bigskip

{\bf JOHN P. RALSTON}\\

\bigskip
{\it Department of Physics and Astronomy\\
University of Kansas\\
Lawrence, KS 66045}\\

\vglue 0.4 in
\ecn
\medskip

\begin{abstract}
{\tenrm\baselineskip=12pt
We review the theoretical formalism for hard exclusive processes in a nuclear
medium.
Theory suggests that these processes will show the very interesting phenomena
of
 color transparency and nuclear filtering. The survival probability in nuclear
media has also been predicted to show a scaling behavior at large momentum
and large nuclear number. We show that all of these effects may have already
been seen experimentally.}
\end{abstract}

\vfill
\eject
\sect{1} {Introduction}

Hard QCD processes in a nuclear medium are expected to show the very
interesting phenomena of color transparency$^1$ and nuclear filtering$^{2,3}$.
Both of these have been suggested by qualitative theoretical
arguements. Partial theoretical proofs based on
factorization in the nuclear medium$^3$ indicate that both phenomena are
important for nuclear number $A>>1$.

Here we report on a systematic study of color transparency
and nuclear filtering. We review the
theoretical formalism which allows us to separate quasi--exclusive nuclear
cross-sections in terms of a perturbative part,
non-perturbative wave functions, and their evolution through the nuclear
medium. This framework can then be used for
phenomenological applications.

An important result is that the usual short distance
formalism for exclusive processes
may be more applicable to processes in the nuclear medium than to
processes in free space.
The reason for this is nuclear filtering of soft components
of the hadron wave function. Any soft components that may
contribute to free space scattering should become depleted in 
 nuclear scattering since the soft components cannot propagate
through nuclei.
Experimental data available so far seem to support these ideas.

A rather remarkable feature that emerges from our study is the
scaling behavior of the nuclear medium effects for large $A$ and
$Q^2$. The nuclear propagation of the hadron is a non-perturbative problem. It
is found that one does not need to solve this problem in the large $A$ and
large
$Q^2$ limit to check the factorization. Instead, the formalism predicts that
survival probabilities should
be a function only of the ratio $Q^2/A^{1/3}$
in this limit. This is a test of factorization plus filtering.
The functional form of this dependence is
theoretically much more model dependent.

\sect {2} {Theoretical Formalism}

By assuming factorization$^3$
the amplitude $M$ can be written in the light cone
gauge as

\beq
M(Q^2, A) = \bigg\lbrace\prod_{i,f}\int [dx d^2k_T]
\bigg[\psi_A^{(f)*}(x_f,k_{Tf})
H(x_i,x_f,k_{Ti},k_{Tf},Q^2)\psi_A^{(i)}(x_i,k_{Ti})\bigg]\bigg\rbrace
\eeq
where $H(x,k_T,Q^2)$ represents the hard scattering,
$\psi_A(x,k_{T})$ represents the hadronic wave function inside the
nuclear medium
and $Q^2$ is the characteristic momentum scale
in the collision.  In this equation we are emphasising the internal quark
coordinates
and suppressing integration over center of mass hadron variables, indicated by
curly brackets.
The wave function represents the overlap
of the short distance hadronic wave function with its propagation
through the nuclear medium and can be expressed as

\beq
\psi_A(x',{k'}_T^2) = \bigg\lbrace\int dx d^2k_T\bigg(
\delta(x-x')\delta^2(k_T-k_T')-
F_A\bigg[s,\bigg({x\over x'}(k_T-k_T')\bigg)^2
\bigg]\bigg)\psi_0(x,k_T^2)\bigg\rbrace
\eeq
where $\psi_0(x,k_T^2)$ is the free space hadronic wave function and
$F_A$ is the nuclear
scattering amplitude. For simplicity here we suppress the $i,f$ indices. The
nuclear filtering
of soft components of the wave function will require that only the
short distance part of the wave function survives. Therefore
the function $\psi_{0A}(x,k_T^2)$ may be approximated very well
by its short distance form, in contrast to the analogous quantity in
free space. Going to the quark transverse separation $(b)$ space we get:
\beq
\tilde \psi_A(x,b) = \tilde f_A(s,x,b)\tilde \psi_0(x,b) ,
\eeq
where the tilde's denote the Fourier transforms and $\tilde f_A=1-\tilde
F_A$ is the nuclear survival amplitude. The distribution amplitude
can then be written as
\beqn
\phi_A(x,Q^2) &=& \int^{Q}_0d^2k_T\int d^2b_T e^{i\bf b_T\cdot\bf k_T}
\tilde f_A(s,x,b^2)\tilde\psi_0(x,b);\nonumber\\
&=&(2\pi)^2 Q\int^{\infty}_0 db J_1(Qb)
\tilde f_A(s,x,b^2)\tilde\psi_0(x,b).
\eeqn
The above integral will get its dominant contribution from the $b\lsim 1/Q$
region of the wave function.
We assume that the wave function $\psi$ is a slowly varying
function of $b$ for small $b$. This assumption is supported by
renormalization group studies. We can
then approximate the above integral by replacing the wave function
by its value at $b\approx 1/Q$ and cutting off the integral at
the same value of $b$. By substituting this expression into Eq. (1)
we see that we have a factorized form for the amplitude $M$:
\beqn
M(Q^2,A)&=& \int [dx] \bigg\lbrace\tilde f_A(b\approx 1/Q)
 \bigg[\psi_{A0}^{(1)}(x_1,1/Q)
H(x,0,Q^2)
\psi_{A0}^{(2)}(x_2,1/Q)\bigg]\bigg\rbrace,\nonumber\\
&\equiv&<\bigg\lbrace\tilde f_A(b^i\approx 1/Q)\ldots\tilde
f_A(b^f\approx 1/Q)\bigg\rbrace H(x,0,Q^2)>,
\eeqn
where the pointed and curly brackets indicate restoration of the integrals
over $x$ and
hadron center of mass coordinates, respectively, which code information on
nuclear size and density. Note that $H(x,0,Q^2)$ is independent of $A$ and
hadron center of mass coordinates for $A>>1$.
The transparency ratio T can then be written as
\beqn
T(Q^2,A) &=& {d\sigma/dt|_A\over A\ d\sigma/dt|_{free\ space}};\nonumber\\
&\cong& {|<\lbrace\tilde f_A(b^i\approx 1/Q)\ldots \tilde f_A(b^f\approx
1/Q)\rbrace H(x,0,Q^2)>|^2\over
Ad\sigma/dt_{free\ space}}\ ;\nonumber\\
&\to& P(Q^2,A) R(Q^2),
\eeqn
where $N$ is the number of participating particles that cross the nucleus.
The transparency ratio can thus be factorized into two pieces:
the survival probability $P(Q^2,A)$ and the ratios of the hard scatterings
$R(Q^2)$.
The leading $Q^2$ dependence of the nuclear hard scattering
can be obtained by considering the leading order perturbation theory
diagrams convoluted with the short distance hadronic wave function.

The above procedure is applicable to all exclusive processes inside
nuclei.
An interesting
prediction of the formalism is that
 the exclusive processes inside nuclear media will get contribution only from
the hard components of the hadronic wave function. Free
space exclusive processes show many phenomena, for example
oscillations in $pp\rightarrow pp$ scattering, helicity violation
in elastic hadron-hadron collisions etc., which give
indisputable evidence of deviations from
the short distance pQCD model. In the nuclear medium, however,
we expect that the short distance model will be applicable and
should accurately predict the hard cross-sections. Therefore
the oscillations in pp elastic
scattering should be absent in experiments on large nuclei. Moreover, since
only the short distance
part of the hadronic wave function is involved inside the
nuclear medium, these processes should show color transparency,
namely that the effective attenuation cross section goes like
$1/Q^2$ for large $Q$. Finally the survival probability part of
the nuclear cross section should depend only on the
variable $Q^2/A^{1/3}$ for large $Q$ and large $A$. As we discuss
in the next section all of these predictions are nicely confirmed
by the BNL experiment on color transparency.

\sect{3} {Experimental Confirmation}

We next show that the BNL experiment$^4$ is consistent with all of the tests
considered
above. The fact that the
oscillations are absent in the nuclear cross section was
pointed out in Ref. (3) immediately after the data was
published. However, since this indicated that the relation of the free space
to nuclear hard scattering
was more subtle than previously assumed, it was not clear how to separate the
survival probability from the (unknown) nuclear hard scattering rate.

The fact that the BNL experiment also shows color transparency
was understood only recently$^5$ with the introduction of a new
method to analyze such experiments. The method is based on Eq. (6) and looking
at the $A$ dependence of the data at fixed $Q^2$. If the nucleus is becoming
more transparent at larger $Q^2$ then the {\it curvature} of the $A$
dependence will become smaller, approaching zero in the limit of
perfect transparency. This trick allows an attenuation cross section to be
extracted
without biasing the analysis with any model for the ratio $R(Q^2)$.
As shown in Ref. (5) the curvature with $A$ indeed
decreases in going from $Q^2=4.8$ GeV$^2$ to $Q^2=8.5$ GeV$^2$.
The effective attenuation cross section $\sigma_{eff}$ extracted from the data
goes roughly like $\sigma_{eff}=40$ mb 2.2 GeV$^2/Q^2$. The
new data analysis procedure also allows independent extraction
of the ratio $R(Q^2)$ of the hard scattering cross sections as defined in Eq.
(6). The hard scattering
cross section in large nuclei was found to follow rather closely the prediction
of the short distance model, again confirming the prediction
that the nucleus filters out most of the soft components.

The scaling law for the survival probability was predicted in Ref. (6).
However at that time the relation between the survival probability and the
transparency ratio was unclear. The theoretical formalism suggests that the
transparency ratio is equal to a function of $Q^2$ times the
survival probability. Given enough data this function as well as
the survival probability can be extracted from data without
any further theoretical input. However at present we have limited
data and to check the scaling law we use
the short distance model for the nuclear hard scattering:
$d\sigma^{hard}/dt\sim \alpha_s(Q^2)^{10}s^{-10}$.
This allows us to
extract the survival probability. The survival
probability extracted in this fashion is found to to be a function of
$Q^2/A^{1/3}$, confirming the scaling law.$^7$ This is an independent
successful
test: it is shown in Fig. (1).

\bigskip

\noindent
{\bf Acknowledgements:}    This work has been supported in part by the DOE
Grants No.
DE-FG02-85-ER-40214.A008 and DE-FG05-91ER-40636.

\bigskip

\noindent {\bf References}
\bigskip
\begin{enumerate}
\item S. J. Brodsky and A. H. Mueller, Phys. Lett. B {\bf 206}, 685 (1988), and
references therein.

\item G. Bertsch, S. Brodsky, A. Goldhaber and J. Gunion, Phys. Rev. Lett. {\bf
47}, 297 (1981).

\item J. P. Ralston and B. Pire, Phys. Rev. Lett. {\bf 61}, 1823 (1988);
ibid {\bf 65}, 2343 (1990).

\item  A. S. Carroll et al., Phys. Rev. Lett. {\bf 61}, 1698 (1988).

\item P. Jain and J. P. Ralston, Phys. Rev. {\bf D48}, 1104 (1993). See also
ibid, {\bf D46}, 3807 (1992).

\item B. Pire  and J. P. Ralston, Phys. Lett. {\bf B256}, 523 (1991).

\item P. Jain and J.P. Ralston, ``Evidence for Observation of Color
Transparency in $pA$ Collisions with Global Fit and Scaling Law Analysis,"
in {\it Proceedings of the XXVIII International Rencontre de
Moriond: QCD and High Energy Interactions}, (Les Arcs 1993) ed. by J. Tran
Thanh Van (Editions Frontiers, in
press).
\end{enumerate}
\vfill
\eject
\null
\vfill
{\tenrm\baselineskip=12pt
Fig. 1. Scaling dependence. The survival probability/constant as a
function of the variable $Q^2/A^\alpha$ for the BNL data for three different
values of $\alpha$. One universal curve could fit all the data for
$\alpha=1/3$.}
\end{document}